\documentclass[12pt]{article}
\usepackage{color} 
\usepackage{amsmath}
\usepackage{amssymb} 
\usepackage{bm}
\usepackage[dvipdfm]{graphicx}
 
\begin{document}
\begin{flushright}
KEK-TH-1561 \\
OIQP-12-07
%{\tt hep-ph/08mmxxx}\\
\end{flushright}
\vspace*{1.0cm}

\begin{center}
\baselineskip 20pt 
{\Large\bf 
Radiation Reaction by Massive Particles\\ 
and Its Non-Analytic Behavior
} 
\vspace{1cm}

{\large 
Satoshi Iso$^{1,2}$\footnote{E-mail address: satoshi.iso@kek.jp},
Sen Zhang$^{3}$\footnote{E-mail address: lightondust@gmail.com},
 \\
} \vspace{.5cm}

{\baselineskip 20pt \it
$^{1}$ Institute of Particle and Nuclear Studies, \\ High Energy Accelerator Research Organization(KEK) \\
 $^{2}$Department of Particles and Nuclear Physics, \\
 The Graduate University for Advanced Studies (SOKENDAI), 
  \\
Oho 1-1, Tsukuba, Ibaraki 305-0801, Japan \vspace{3mm} \\
$^{3}$ Okayama Institute for Quantum Physics, \\
Kyoyama 1-9-1, Kita-ku, Okayama 700-0015, Japan
%\vspace{3mm}
}

\vspace{2cm} 
% }}}
{\bf Abstract} %{{{
\end{center}
\noindent
We derive a massive analog of the ALD (Abraham, Lorentz and Dirac) equation, i.e., the equation
of motion of a  relativistic charged particle with a radiation reaction term
induced by emissions of massive fields.
We show that the radiation reaction term 
 has a non-analytic behavior as a function of the mass $M$ of the radiation field and
both  expansions with respect of $M$ and $1/M$ are generally invalid.
Hence the massive ALD equation cannot be written as a local equation with derivative expansions.
We especially investigate the radiation reaction in
three specific motions,
uniform acceleration, a circular motion and a scattering process.

\thispagestyle{empty}
 
%\bigskip
\newpage

\addtocounter{page}{-1}

%%%%%%%%%%%%%%%%%%%%%%%%%%
%\baselineskip 36pt
% Main body
%%%%%%%%%%%%%%%%%%%%%%%%%%
\baselineskip 18pt
%%%%%%%%%%%%%%%%%%%%%%%%%% }}}
\section{Introduction} %{{{

A charged particle emits radiation when it is accelerated
and loses the kinetic energy. Effectively the emission induces a friction term in the 
equation of motion of the charged particle. 
The modified equation with the radiation reaction term becomes
\begin{eqnarray}
m \ddot{z}^\mu = F^\mu_{ext} + \alpha (\dddot{z}^\mu + \dot{z}^\mu \ddot{z}^\nu \ddot{z}_\nu)
\end{eqnarray}
and it is called the ALD equation \cite{ALD}. 
The first term is the external force while the second one
corresponds to the back-reaction of the radiation 
of electromagnetic fields.
The coefficient $\alpha$ of the radiation reaction force is
$\alpha = e^2/6\pi$ when the massless photon is radiated.
The ALD equation itself  is well established
but it has an infamous problem of the runaway solutions.
The ALD equation contains the third derivative term and
 the point particle is {\it accelerated} to the speed of light by the radiation force.
One can remove such pathological solutions 
by imposing a regular boundary condition at the infinite future.
But then, the solutions must be accelerated before the external force is applied.
It is against the causality and called the problem of the preacceleration.
A pragmatic resolution is to treat the backreaction term as a perturbation \cite{LL},
and replace the ALD equation by the so called  Landau Lifshitz equation.
The runaway solution and the problem of preacceleration  can  then be removed but the
validity of the Landau-Lifshitz equation is restricted 
to  a situation  when the acceleration is sufficiently small.
It is also important to  investigate how  the 
correction to the ALD equation is obtained from
 the quantum field theory calculation in which
 there is no such pathological problems
(see, e.g. \cite{Higuchi:2005gh,Higuchi:2005an,Higuchi:2006xk,Johnson:2000qd}).

An important aspect of the ALD equation can also be inferred from the study of
 the radiation reaction in other situations such as 
in other space-time dimensions \cite{Shuryak:2011tt},  in  curved backgrounds
or the backreaction due to the gravitational radiation \cite{DeWitt:1960fc,Mino:1996nk,Quinn:1996am}.
In all of these situations 
the radiation reaction term becomes  nonlocal,
and a local differential equation such as the ALD
equation can be obtained only in special limits.  
Hence it will be important to see whether such a nonlocal effect   also arises 
when the radiation field becomes massive, and to see 
how such a nonlocal effect is related to
the problem of runaway solutions.

In this paper, we consider a relativistic point particle coupled with a massive scalar field, and
investigate properties of radiation reaction by emissions of the massive scalar fields. 
This system is considered as a toy model of the ALD equation with massive photons in plasma.
We derive an analog of the ALD equation, but the radiation reaction term becomes generally nonlocal 
unlike the massless ALD equation, and the derivative expansion ($1/M$ expansion)
can not be applied.  
Furthermore, 
we show that the massless limit $M \rightarrow 0$ is nonanalytic: the 
coefficient of the radiation reaction 
term contains a logarithmic term like $M^2 \log M$ or $M^4 \log M$.

The paper is organized as follows.
In section 2, we derive the massive analog of the ALD equation.
We first develop several tools for the calculation and 
 compare the radiation reaction in massless and  massive cases. 
The backreaction (radiation reaction) 
term in the massive case behaves non-analytically as a function of the mass $M$
and contains a logarithmic term like $M^2 \log M$.
We also show that 
 the derivative expansion (namely an expansion with respect to $1/M$) is generally invalid.
Since the term itself is not singular as a function of $M$, it shows that the radiation reaction term 
is essentially nonlocal. 
In section 3, in order to see the explicit behavior of the backreaction,
we evaluate the radiation reaction term numerically for specific motions, uniform acceleration and a circular motion.
We show that, in the uniform acceleration, 
the radiation reaction term behaves like $M^2 \log{M}$ near $M=0$.
We also discuss a scattering process (a motion when the external force is applied
during a finite time interval) to see that the nonanalytic behavior is not specific to the uniformly accelerated
motion.
In appendix \ref{Bessel}, we discuss the non-analytic behavior of the backreaction based on the 
properties of the Bessel function which appears in the propagator of the massive scalar fields.

%%%%%%%%%%%%%%%%%%%%%%%%%%%%%%%%%%%%%%%%%%%%%%%%%%%%%%%%%%
%%%%%%%%%%%%%%%%%%%%%%%%%%%%%%%%%%%%%%%%%%%%%%%%%%%%%%%%%% }}}
\section{Massive ALD Equation and Radiation Reaction} %{{{
\setcounter{equation}{0} 
\subsection{Derivation of the Radiation Reaction Force}
The action of a relativistic charged particle 
interacting with a scalar field is given by
\begin{eqnarray}
S = -m_0 \int d\tau \ \sqrt{\dot{z}^\mu \dot{z}_\mu} + 
\int d^4x \ j(x;z) \phi(x)
+ \int d^4 x \frac{1}{2} ( \partial_\mu \phi(x) \partial^\mu \phi(x) - M^2 \phi^2(x))
\end{eqnarray} 
where the scalar current $j(x;z)$ is defined as
\begin{eqnarray}
 j(x;z) = e \int d\tau \ \sqrt{\dot{z}^\mu \dot{z}_\mu} \delta^4 (x-z(\tau)).
  \label{current}
\end{eqnarray}
In the paper, we set the speed of light $c=1$.
This model is considered as a toy model of a charged point particle in massive photon fields
in plasma.
The equations of motion of the coupled system of  the position of the particle $z(\tau)$ 
and the scalar field $\phi(x)$ are given by
\begin{eqnarray}
& m_0 \ddot{z}^\mu = -e( \partial^\mu - \dot{z}^\mu \dot{z}^\nu \partial_\nu - \ddot{z}^\mu ) \phi(z), 
\label{zeom} \\
& ( \partial^\mu \partial_\mu + m^2 )\phi(x) = j(x;z). \label{phieom}
\end{eqnarray}
The parameter $\tau$ is chosen to satisfy the gauge condition
 $\dot{z}^\mu \dot{z}_\mu = 1$. 
The above equation of motion is consistent with this gauge condition.
In order to derive the massive ALD equation, we
write the scalar field $\phi$ as a sum of an external part and a self-interaction part,
 $\phi(x) = \phi_{ext}(x) + \phi_{self}(x)$.
Then the equation of motion (\ref{zeom}) for $z(\tau)$ becomes
\begin{eqnarray}
m_0 \ddot{z}^\mu &=& F_{ext}^\mu +F_{self}^\mu 
%- e( \partial^\mu - \dot{z}^\mu \dot{z}^\nu \partial_\nu - \ddot{z}^\mu ) \phi_{self}(z),
\end{eqnarray}
%here $\phi_{ext}$ denotes the field due to external source, and it is related to the external force by
where the external force $F_{ext}^\mu $ is given by the external field $\phi_{ext}(x)$ as
\begin{eqnarray}
F_{ext}^\mu (z)= 
-e( \partial^\mu - \dot{z}^\mu \dot{z}^\nu \partial_\nu - \ddot{z}^\mu ) \phi_{ext}(z).
\end{eqnarray}
On the contrary, the self-force (radiation reaction force)  $F_{self}$ is  generated by 
the charged current of the particle itself, and written in terms of  the induced field $\phi_{self}(x)$ as
\begin{eqnarray}
F_{self}^\mu (z)= 
-e( \partial^\mu - \dot{z}^\mu \dot{z}^\nu \partial_\nu - \ddot{z}^\mu ) \phi_{self}(z).
\end{eqnarray}
Here the induced field $\phi_{self}(x)$ is solved
 in terms of  the current $j(x';z)$ 
 by using the retarded Green function
\begin{eqnarray}
\phi_{self}(x) = \int d^4 x' G_R(x,x') j(x';z)
\end{eqnarray}
where $G_R(x,x')$ satisfies the equation
\begin{eqnarray}
( \partial^\mu \partial_\mu + M^2) G_R(x,x') = \delta^{(4)}(x-x').
\end{eqnarray}
Let us  first discuss some general properties of  the radiation reaction force 
without using the explicit form of $G_R(x,x')$.

Due to the Lorentz symmetry, the Green function can be written
as  a function of the distance $\sigma = (x-x')^2$,
\begin{eqnarray}
G_R(x,x') = \theta(x^0- x'^0) G(\sigma).
\end{eqnarray}
Substituting this expression in $\phi_{self}(x)$, one can rewrite the self-force as
\begin{eqnarray}
F_{self}^\mu (z) = 
- e^2 ( \partial^\mu - \dot{z}^\mu \dot{z}^\nu \partial_\nu - \ddot{z}^\mu ) 
\int^{\tau(x^0)}_{-\infty} d\tau' G_R((x-z(\tau'))^2)_{|x=z(\tau)},
\label{xform}
\end{eqnarray}
where $\tau(x^0)$ is defined by the  proper time 
$\tau$ which satisfies $z^0(\tau) = x^0$. 
In taking  derivatives with respect to $x$, 
one first need to evaluate the integral at general $x$, and 
then take the limit of  $x$ to a point $z(\tau)$ on the trajectory of the particle.
But noting that
 $G(\sigma)$ depends on $x^\mu$ and $x'^\mu$ only through their distance
$\sigma$, one can express the right hand side of (\ref{xform}) in terms of  the
quantities on the trajectory,
\begin{eqnarray}
\partial_\mu \int^{\tau(x^0)}_{-\infty} d\tau' \ G(\sigma)_{|x=z(\tau)} = 
\int^{\tau}_{-\infty} d\tau' \ 2(z^\mu(\tau) - z^\mu(\tau')) \frac{d}{d\sigma} G(\sigma).
\label{Gint}
\end{eqnarray}
The integrand is a function of $\sigma =(z(\tau)-z(\tau'))^2$ and its derivatives
while the integral is performed over the proper time $\tau'$ of the trajectory. Since $\sigma$ and 
$s=\tau-\tau'$ are related to each other, we can either express the  integral (\ref{Gint})
in terms of  $\sigma$ or $s$. Either expression has its own advantage, so we will explain both
expressions in the following.
  
We first express the integral (\ref{Gint}) and the self-force term (\ref{xform})
in terms of $s \equiv \tau - \tau'$. First $z(\tau)-z(\tau')$ can be expanded 
in a power series of $s$ as
\begin{eqnarray}
y^\mu(s;\tau) \equiv z(\tau) - z(\tau') = - \sum^{\infty}_{n=1} \frac{(-s)^n z^{(n)\mu}}{n!}
= s \dot{z}^\mu(\tau) - \frac{s^2}{2}\ddot{z}^\mu (\tau) + \frac{s^3}{6} \dddot{z}^\mu (\tau)+ \cdots,
\end{eqnarray}  
where $z^{(n)\mu}(\tau) = d^n z^\mu(\tau)/d\tau^n$. 
We fix $\tau$ and change the variable from $\tau'$ to $s=\tau-\tau'$. 
By using the gauge condition 
$\dot{z}^\mu \dot{z}_\mu = 1$, $\dot{z}^\mu \ddot{z}_\mu = 0$ 
and $\dot{z}^\mu \dddot{z}_\mu + \ddot{z}^\mu \ddot{z}_\mu = 0$,
it is straightforward to check the following relations
\begin{eqnarray}
\sigma(s;\tau) &=& y^\mu y_\mu = s^2 (1- \frac{s^2}{12} \ddot{z}^\mu \ddot{z}_\mu + \cdots), \nonumber \\
\frac{d\sigma(s;\tau)}{ds} &=& 2 y^\mu \dot{y}_\mu = 2s (1 - \frac{s^2}{6} \ddot{z}^\mu \ddot{z}_\mu + \cdots).
\end{eqnarray}
Using the relation $d/d\sigma = (ds/d\sigma) d/ds$, one obtains 
\begin{eqnarray}
F^\mu_{self} = -e^2 \int^{\infty}_{0} ds 
(P^\mu_\nu \frac{y^\nu}{y^\rho \dot{y}_\rho} \frac{d}{ds} - \ddot{z}^\mu(\tau) ) G(\sigma(s;\tau)),
\end{eqnarray}
where $P^\mu_\nu = \delta^\mu_\nu - \dot{z}^\mu(\tau) \dot{z}_\nu(\tau)$ is a projection operator which satisfies $ P^\mu_\nu \dot{z}^\nu (\tau)= 0$.
The surface term 
does not contribute because 
$P^\mu_\nu (y^\nu/y^\rho \dot{y}_\rho) \rightarrow P^\mu_\nu \dot{z}^\nu = 0$ near $s=0$.
So we can perform an integration by parts and get the expression of the self-force term in terms of $s$
\begin{eqnarray}
F^\mu_{self} = e^2 \int^{\infty}_{0} ds \left[
P^\mu_\nu \frac{d}{ds} \bigg(\frac{y^\nu}{y^\rho \dot{y}_\rho} \bigg) + \ddot{z}^\mu(\tau) 
\right] G(\sigma).
\label{sform}
\end{eqnarray}
In order to evaluate it, let us first
 expand the integrand in powers of $s$,
\begin{eqnarray}
F^\mu_{self} &=& e^2 \int^\infty_0 ds 
\left[ P^\mu_\nu \big{\{}\frac{d}{ds}
\big( -\frac{s}{2} \ddot{z}^\mu+ 
\frac{s^2}{6}(\dddot{z}^\mu + \dot{z}^\mu \ddot{z}^\rho\ddot{z}_\rho) + \cdots 
\big)\big{\}}
 + \ddot{z}^\mu \right]
 G(\sigma) \nonumber \\
 &=& e^2 \int^\infty_0 ds \left[
 \frac{1}{2} \ddot{z}^\mu + \frac{s}{3}(\dddot{z}^\mu + \dot{z}^\mu \ddot{z}^\rho \ddot{z}_\rho)
  + \cdots
\right] G(\sigma).
\label{sformexpansion}
\end{eqnarray}
As we will see later, for a massless scalar field
where $G(\sigma) \sim \delta(\sigma)$, 
higher order terms of $s$ vanish except for the mass renormalization
 and the radiation reaction 
term. Thus one obtains the ALD equation, which is written as a local equation.
On the other hand,  for a massive scalar field, $G(\sigma)$ only damps as an 
inverse-power of $\sigma$ at $s \rightarrow \infty$.
Then  integrals of $s$ for sufficiently higher orders  diverge. 
Since the integral (\ref{sform}) is shown to be finite after the mass renormalization,
the divergence of the coefficients for higher orders
indicates that the derivative expansion of the radiation reaction term is not valid.
Hence the massive analog of the ALD equation cannot be written as a sum of local terms. 
We therefore
need to evaluate the integral (\ref{sform}) directly without the derivative expansion, and
for this purpose, it is more convenient to express the integral in terms of $\sigma$ instead of $s$.

The Green function $G(\sigma)$ is generally a 
complicated function of $\sigma$
(e.g. see (\ref{greenfunction}) for a massive scalar field.
Therefore it is sometimes more appropriate to express the integral in terms of $\sigma$
instead of $s$, especially in evaluating the radiation reaction term numerically.
The integral (\ref{sform}) is rewritten by changing the integration variable from $s$ to $\sigma$ as
\begin{eqnarray}
F^\mu_{self} = e^2 \int^{\infty}_{0} d\sigma \left[
P^\mu_\nu \bigg{(} \frac{d}{d\sigma} 2y^\nu \frac{ds}{d\sigma} \bigg) 
+ \ddot{z}^\mu(\tau)\frac{ds}{d\sigma} 
\right] G(\sigma).
\label{sigmaform} 
\end{eqnarray}
In the region $0 < s < \infty$, the function $\sigma(s)$ is single valued 
and  $d\sigma/ds> 0$ is satisfied.
Hence we can solve $s$ as a function of $\sigma$.
By using the variable $l = \sqrt{\sigma}$, the self-force can be expressed as 
\begin{eqnarray}
F^\mu_{self} &=& e^2 \int^{\infty}_{0} dl 
\left[
P^\mu_\nu \bigg{(} \frac{d}{dl} y^\nu \frac{1}{l} \frac{ds}{dl} \bigg) 
+ \ddot{z}^\mu(\tau)\frac{ds}{dl} 
\right] G(l^2) \nonumber \\
&=& e^2 \int^\infty_0 dl 
\left[ \frac{\ddot{z}^\mu}{2} + (\dddot{z}^\mu + \dot{z}^\mu \ddot{z}^\nu \ddot{z}_\nu )\frac{l}{3} + \cdots
\right] G(l^2).
\label{sigmaformexpansion}
\end{eqnarray}
In the second equality, we have used the following relations
\begin{eqnarray}
\frac{ds}{dl} &=& 1 + \frac{\ddot{z}^\nu \ddot{z}_\nu}{4} \frac{l^2}{2} + o(l^3),
\nonumber \\
y_{\mu} &=& \dot{z}_\mu l - \ddot{z}_\mu \frac{l^2}{2} + 
\left(\dddot{z}_\mu + \frac{\dot{z}_\mu \ddot{z}^\nu \ddot{z}_\nu}{4} \right)
\frac{l^3}{6} + o(l^4).
\end{eqnarray}
The usefulness of 
the expression (\ref{sigmaformexpansion}) becomes apparent when we consider, e.g. the renormalization 
of  the mass term $m \ddot{z}$. 
The radiation reaction term, (\ref{sform}) or (\ref{sigmaform}), contains 
a divergence which is  proportional to $\ddot{z}^\mu$, and should be absorbed by the mass renormalization.
For a massless scalar field with $G(\sigma) \sim \delta(\sigma) = \delta(s^2)$, it makes   no difference
whether we subtract the divergent term  in  (\ref{sform}) or (\ref{sigmaform}).
However, for a massive case, $G(\sigma)$ is a nontrivial function of 
$\sigma(z,z') = (z^\mu(\tau) - z^\mu(\tau'))^2$, and
since $\sigma$ depends on $z^\mu$ and $\tau'$ in a complicated way, 
an integral of $G(\sigma)$ such as  $\int ds f(s) \ddot{z}^\mu(\tau) G(\sigma)$ depends 
not only on $\ddot{z}^\mu(\tau)$ but also on the details of the trajectory $z(\tau')$.
On the other hand, an integral of $G(\sigma)$  such as
 $\int d\sigma f(\sigma) \ddot{z}^\mu(\tau) G(\sigma)$  is a  constant multiplied by
 $\ddot{z}^\mu(\tau)$. 
Therefore the mass renormalization of the massive ALD equation should be evaluated
not 
in the expression (\ref{sform}) but in (\ref{sigmaform}).
 
%%%%%%%%%%%%%%%%%%%%%%%%%%%%%%%%%%%%%%%%%%%%%%%%%%%%%%%%%% }}}
\subsection{Massless Radiation} %{{{
Before considering a massive case, we briefly review the evaluation of the 
radiation reaction by a massless scalar field.
For a massless scalar field, the retarded Green function is simply given by
\begin{eqnarray}
G_R(x-x') = \theta(x^0-x'^0) \frac{\delta(\sigma)}{2\pi}.
\end{eqnarray}
By substituting it in (\ref{sformexpansion}) or (\ref{sigmaformexpansion})
\footnote{The integration seems subtle 
since the delta function is just at the edge of the integration range. But 
the subtlety can be removed by going back to the original equation (\ref{xform}).
Here the integral is performed over the region $\tau(x^0)> \tau'> - \infty$ 
and  the delta function is always inside the region. },
 we can obtain the ALD equation
\begin{eqnarray}
m \ddot{z}^\mu = F^\mu_{ext} + \frac{e^2}{12\pi} (\dddot{z}^\mu + \dot{z}^\mu \ddot{z}^\mu \ddot{z}_\mu).
\end{eqnarray}
Here the divergent term
\begin{eqnarray}
e^2 \int^\infty_0 dl \ \ddot{z}^\mu \frac{\delta(l)}{4l}
\end{eqnarray}
is absorbed by the mass renormalization.

We would like to mention the derivation of the ALD equation 
 from the quantum field theory calculations \cite{Higuchi:2006xk}.
The authors evaluated two types of diagrams which contribute to the backreaction.
One type of diagrams corresponds to the radiation processes.
The contribution of this type to the backreaction turns out to be written 
in the form of (\ref{xform}) but with $G_R$ replaced by $G_-$
\begin{eqnarray}
G_-(x,x') = G_R(x,x') - G_A(x,x') = \frac{\theta(x^0-x'^0) - \theta(x'^0-x^0)}{2} G(\sigma),
\end{eqnarray}
where $G_A(x,x')$ is the advanced Green function. 
With this replacement, the terms with even powers of $s$ in (\ref{sformexpansion}) vanish 
and the remaining terms become finite.
Another type of diagrams corresponds to the self-energy (including the mass renormalization)
of the point particle.
It has a form of (\ref{xform}) but with $G_R$ replaced by $G_+$
\begin{eqnarray}
G_+(x,x') = G_R(x,x') + G_A(x,x').
\end{eqnarray}
It gives a divergent term which can be  absorbed by the mass counter term.
Summing these two types of diagrams,
 one can reproduce the ALD equation
\footnote{One might expect to subtract a divergence
by replacing $G_R$ by $G_-$ or $G_F$ also for the massive case discussed in the next subsection.
But the Green function in the massless case depends  on $\sigma$ only through $\delta(\sigma)$, 
and the acausal time dependence of $G_A$ inside $G_-$ and $G_F$ does not appear in the final results. 
However, in a massive case, the Green function has a  tail and
a simple replacing of $G_R$ by another type of Green functions such as $G_\pm$
violates the causality explicitly.}.
  
%%%%%%%%%%%%%%%%%%%%%%%%%%%%%%%%%%%%%%%%%%%%%%%%%%%%%%%%%% }}}
\subsection{Massive Radiation} %{{{
\label{massive}
The retarded Green function of a massive scalar field is given by
\begin{eqnarray}
G_R(x,x') = \frac{\theta(x^0-x'^0)}{4\pi} \left( 
2\delta(\sigma) - \theta(\sigma) \frac{M J_1(M \sqrt{\sigma})}{\sqrt{\sigma}}
\right).
\label{greenfunction}
\end{eqnarray}
The first term in the parentheses is the same as the massless case 
while the second term represents a modification of the Green function 
by the  mass of the radiation field.
Since the second term disappears in the massless limit, one might expect that 
the massive ALD equation can be written as a perturbation of the massless ALD equation by
higher derivative terms. The situation is not so simple, however.
By the dimensional analysis,
the coefficient of the derivative expansion must contain an inverse power of $M$, 
but 
such an expansion is shown to be invalid because of the
nonanalyticity at $M=0$.

To see such nonanalytic behaviors of the derivative expansion, let us try to evaluate each coefficient of 
the derivative expansion by
 substituting the Green function (\ref{greenfunction}) in (\ref{sigmaformexpansion}).
The leading order term is proportional to $\ddot{z}^\mu$ which corresponds to the mass renormalization.
The next order term gives the coefficient of the ALD term. 
The coefficient, however,  vanishes
due to the cancellation between the massless propagator $\delta(\sigma)$ and the 
massive modification of the Green function,
\begin{eqnarray}
e^2 (\dddot{z}^\mu + \dot{z}^\mu \ddot{z}^\nu \ddot{z}_\nu) \int^\infty_0 dl 
\frac{l}{12\pi} \left( 
2\delta(l^2) - \frac{M J_1(M l)}{l}
\right) = 0.
\end{eqnarray}
This cancellation  is universal and does not depend on the value of $M$.
We are tempted to conclude that the ALD term vanishes for a massive scalar field, 
but it is not correct since the derivative expansion is shown to be invalid.
It suggests that something strange happens in the massless limit. 
Let us evaluate the coefficients of higher order terms.
They are written as
\begin{eqnarray}
e^2 \int^\infty_0 dl \ l^{n+1} \frac{M J_1(Ml)}{l}.
\end{eqnarray}
Since the Bessel function behaves as $J_1(x) \sim 1/\sqrt{x}$ for large $x$ (see Appendix A), 
the integral  diverges for $n>1/2$.
Therefore the derivative expansion (or equivalently a power-series expansion with respect to $l=\sqrt{\sigma}$)
is not justified. 
One can show that the expansion in terms of $s$ in (\ref{sformexpansion}) is also divergent.
Such singular behavior of the derivative expansion in the massive case shows a peculiarity of 
the massless ALD equation where the radiation reaction force is written as a local derivative term.

Though one can not write the radiation reaction term in terms of local derivative terms,
the backreaction itself is finite even in the massive case.
To confirm the finiteness of the backreaction, we go back to (\ref{sigmaform}) or (\ref{sform}).
The integration can be divided into two parts, one involving the  integration of $\delta(\sigma)$ and 
the other involving the integration of $J_1(M\sqrt{\sigma})$.
Since the first term is the same as the massless case, it is sufficient to show the finiteness of the 
integrals of the second part,
\begin{eqnarray}
\int^\infty_0 d\sigma \ 
P^\mu_\nu \frac{d}{d\sigma} \left(
 y^\nu \frac{ds}{d\sigma} 
\right)
\frac{ M J_1(M\sqrt{\sigma})}{\sqrt{\sigma}},
\label{nonlocalterm}
\end{eqnarray}
and
\begin{eqnarray}
\int^\infty_0 d\sigma \ 
\frac{ds}{d\sigma} \frac{ M J_1(M\sqrt{\sigma})}{\sqrt{\sigma}}.
\label{nonlocaltermmass}
\end{eqnarray}
Note that for general trajectories 
$P^\mu_\nu (d/d\sigma) ( y^\nu ds/d\sigma )$ or $ds/d\sigma$
is regular at finite $\sigma$ and falls off faster than $\sigma^{-1/2}$
at $\sigma \rightarrow \infty$(or $s \rightarrow \infty$). 
On the other hand, $M J_1(M\sqrt{\sigma})/\sqrt{\sigma}$ is regular at finite $\sigma$ and behaves like
\begin{eqnarray}
\frac{1}{\sigma} \cos{\sqrt{\sigma}}
\end{eqnarray}
at infinity.
So the integration (\ref{sigmaform}) is finite for a fixed $M$, and we get a finite amount of 
backreaction by emissions of a massive scalar field as expected.
We see this finiteness explicitly in the next section for special types of trajectories.

We have seen that, if we perform the derivative expansion,
the coefficient of the ALD term vanishes for a finite $M$ while it becomes $(e^2/12\pi)$ at $M=0$.
One may  also be interested in the mass expansion near $M = 0$ of the backreaction 
term and how
the local equation can be derived from the  massive ALD equation
with a nonlocal radiation reaction term,  (\ref{sigmaform}) or (\ref{sform}).
But it is not so simple as one might expect.
For example, consider the convergent integral (\ref{nonlocalterm}).
If we naively expand it as a power series of $M$, we get
\begin{eqnarray}
M^2 \int^\infty_0 \ 
P^\mu_\nu \frac{d}{d\sigma} \left(
 y^\nu \frac{ds}{d\sigma} 
\right) \sum^\infty_{n=0} \frac{(-1)^n (M\sqrt{\sigma}/2)^{2n}}{N!\Gamma(n+2)}.
\end{eqnarray}
But the higher order terms of $M$ are divergent, and such an expansion is not valid. 
The backreaction must have a non-analytic behavior at $M=0$.
We will investigate this kind of non-analyticity  in the next section.

%%%%%%%%%%%%%%%%%%%%%%%%%%%%%%%%%%%%%%%%%%%%%%%%%%%%%%%%%%%%%
\subsection{Mass Renormalization}
In this subsection, we  discuss  the  mass renormalization.
Because of the non-locality of  the backreaction term, the mass renormalization becomes 
more subtle than the massless case.
Since the non-local term is convergent, one may naively renormalize the mass only
by the divergent integral. But then the renormalization of mass becomes independent of $M$
and survives even in the infinite $M$ limit where  the emission of such a massive field is highly suppressed.
More explicitly, because of the identity of the Bessel function,
\begin{eqnarray}
\lim_{M\rightarrow\infty} M \theta(\sigma) \frac{J_1(M\sqrt{\sigma})}{\sqrt{\sigma}} = 2 \delta(\sigma),
\label{limitJ} 
\end{eqnarray}
the Green function vanishes in the large $M$ limit
\begin{eqnarray}
\lim_{M\rightarrow \infty} G_R(x,x') &=& \lim_{M\rightarrow\infty} \frac{\theta(x^0-x'^0)}{4\pi} \left(
2\delta(\sigma) - \theta(\sigma) \frac{M J_1(M\sqrt{\sigma})}{\sqrt{\sigma}} 
\right) = 0.
\end{eqnarray}
The backreaction therefore must vanish in the limit $M \rightarrow \infty$.
Thus  we also need to take into account the non-local but convergent integrals, 
(\ref{nonlocalterm}) and (\ref{nonlocaltermmass}), in discussing the mass renormalization at finite $M$.

In order to make a consistent mass renormalization, we first introduce a function $f(l)$ and define the 
renormalized mass term by 
\begin{eqnarray}
\delta m \ \ddot{z}^\mu = - e^2 \int^\infty_0 dl \ \ddot{z}^\mu f(l) G(l^2).
\label{massrenorm}
\end{eqnarray}
Then the self-force term can be divided into the mass renormalization $\delta m \ddot{z}^\mu$
and the radiation reaction force $F\mu_{0,self}$ as
\begin{eqnarray}
F_{self}^\mu &=& F_{0,self}^\mu + \delta m \ \ddot{z}^\mu.
\end{eqnarray}
The function $f(l)$ must satisfy the condition $f(0)=1/2$ so that the divergent term is 
correctly subtracted in  (\ref{sform}).
It is consistent with the mass renormalization in the massless case.
For the massive scalar field, the above subtraction  (\ref{massrenorm}) has a contribution 
not only from the term $\delta(\sigma)$ in the Green function but also
from the non-local term $\sim J_1(M\sqrt{\sigma})/\sqrt{\sigma}$.
Another condition for $f(l)$ is that the mass renormalization should vanish in the large $M$ limit.
The simplest choice is $f(l)=1/2$, but the choice is not  unique.
The ambiguous part of the mass renormalization is  written as
\begin{eqnarray}
\delta m_{n} = -e^2 \int^\infty_0 dl f(l) \frac{M J(M l)}{l}.
\end{eqnarray}
It is independent of the trajectory of the  particle.
Hence a different choice of $f(l)$ does not 
affect how the backreaction term depends on the details of the trajectory 
such as the acceleration or the velocity of the particle's motion. 
%The component of the backreaction on $\dddot{z}^\mu + \dot{z}^\mu \ddot{z}^\nu \ddot{z}_\nu$ is also $f(l)$ independent.
In this paper, we take the simplest choice $f(l)=1/2$.
In the next section, we will show, 
if we renormalize the mass by $f(l)=1/2$,
the radiation reaction term becomes $o(a^3)$ 
when a particle is uniformly accelerated with an acceleration $a$.
%\begin{eqnarray}
%\textcolor{red}{ 
%F_{ext}^\mu = m \ddot{z}^\mu + o(a^3),
%}
%\label{massdefinition}
%\end{eqnarray}
For other choices, it is generally proportional to $o(a)$.
This seems to justify the simplest choice $f(l)=1/2$.

%%%%%%%%%%%%%%%%%%%%%%%%%%%%%%%%%%%%%%%%%%%%%%%%%%%%%%%%%
%%%%%%%%%%%%%%%%%%%%%%%%%%%%%%%%%%%%%%%%%%%%%%%%%%%%%%%%%% }}}
\section{Radiation Reaction for Specific Trajectories} %{{{
\setcounter{equation}{0} 
In order to investigate the non-analytic properties of the backreaction, 
we evaluate (\ref{sform}) and (\ref{sigmaform}) explicitly for some specific trajectories.
We consider three cases, a uniform acceleration, a circular motion and a scattering process.
%%%%%%%%%%%%%%%%%%%%%%%%%%%%%%%%%%%%%%%%%%%%%%%%%%%%%%%%%% }}}
\subsection{Uniform Acceleration} %{{{
\label{uniformlyacceleration}
 
A uniformly accelerated point particle is described by the trajectory
\begin{eqnarray}
z^\mu = (\frac{1}{a} \sinh{a \tau},\frac{1}{a} \cosh{a\tau},0,0),
\label{uniftraj}
\end{eqnarray}
where $a$ is the acceleration and $\ddot{z}^\nu \ddot{z}_\nu = -a^2$.
$\dot{z}^\mu(\tau)$ and $\ddot{z}^\mu(\tau)$ 
 form a two dimensional vector space
and other derivatives are written in terms of them as follows
\begin{eqnarray}
z^{(2n)\mu} = a^{2(n-1)} \ddot{z}^\mu, \ \ \ \ \ 
z^{(2n+1)\mu} = a^{2n} \dot{z}^\mu, \ \ \ \ \ 
\ddot{z}^\mu \dot{z}_\mu = 0.
\end{eqnarray}
For the trajectory (\ref{uniftraj}), the radiation reaction term vanishes
\begin{eqnarray}
\dddot{z}^\mu + \dot{z}^\mu \ddot{z}^\nu \ddot{z}_\nu = 0.
\label{rrvanish}
\end{eqnarray}
$l(s)$ or $\sigma(s)$ has the following simple form 
\begin{eqnarray}
l = \sqrt{\sigma} = \sqrt{(z^\mu(\tau) - z^\mu(\tau'))(z_\mu(\tau) - z_\mu(\tau'))} = \frac{2}{a} \sinh{\frac{a(\tau-\tau')}{2}}.
\end{eqnarray}
The backreaction  becomes proportional to 
$\ddot{z}^\mu$ since $P^\mu_\nu \dot{z}^\nu=0$
and is given by
\begin{eqnarray}
F_{self}^{\mu} = \frac{\ddot{z}^\mu}{a} F(a,M) =
e^2 a (\sinh{a\tau},\cosh{a\tau},0,0) \int^{\infty}_0 d l \  
\left(
\frac{1+a^2 l^2/2}{2(\sqrt{1+a^2 l^2 /4})^3} - \frac{1}{2}
\right) G_R(l) .
\end{eqnarray}
Because of (\ref{rrvanish})
the delta function term in (\ref{greenfunction}) vanishes in $F_{self}^\mu$. Then
$F(a,M)$ becomes 
\begin{eqnarray}
F(a,M) &=& - e^2 a \int^\infty_0 dl \left(
\frac{1+a^2 l^2/2}{2(\sqrt{1+a^2 l^2 /4})^3} - \frac{1}{2}
\right) \frac{M J_1(M l)}{4\pi l} \nonumber \\
&=& - e^2 a M \int^\infty_0 d t \left(
\frac{1+ 2 g t^2}{2(\sqrt{1+ g t^2})^3} - \frac{1}{2}
\right) \frac{J_1(t)}{4\pi t} \nonumber \\
&\equiv & e^2 a M \bar{F}(g),
\label{uniformmaster}
\end{eqnarray}
where $g = a^2/4M^2$. In the second equality, we changed the integration variable 
from $l$ to $t=M l$.
All information of the backreaction to the uniformly accelerated particle is contained in
the function $F(a,M)$ or equivalently in $\bar{F}(g)$.

\begin{figure}
\centering
\includegraphics[scale=0.8]{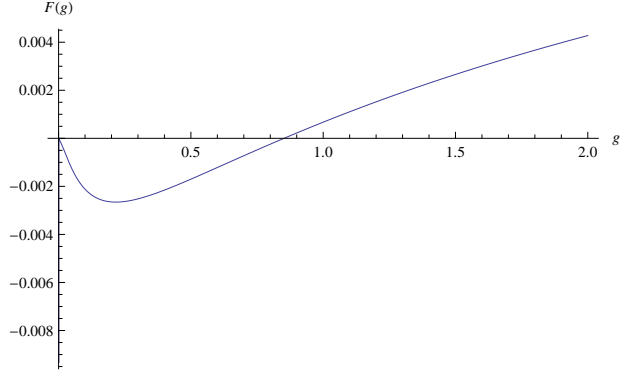}
\caption{The function $\bar{F}(g)$ is plotted as a function of $g= a^2/4M^2$. }
\label{uniformg}
%Figure. \ref{uniformg} 
\end{figure}

The function $\bar{F}(g)$ can be expressed by using the Meijer's G-function (see for example \cite{Meijer})
\begin{eqnarray}
\bar{F}(g) &=& \frac{1}{8\pi} \left( 1 - \int^\infty_0 d t \frac{1 + 2g t^2}{(\sqrt{1+gt^2})^3} \frac{J_1(t)}{t}
\right) \nonumber \\
&=& \frac{1}{8\pi} - \frac{I_1(\frac{1}{2\sqrt{g}}) K_1(\frac{1}{2\sqrt{g}})}{8\pi \sqrt{g}} +
\frac{g}{\pi^{3/2}}	G^{21}_{13}\left( \frac{1}{4g} 
\big|^{1}_{3/2, 3/2, 1/2}
\right).
\end{eqnarray} 
The behavior of the function $\bar{F}(g)$ is plotted in figure. \ref{uniformg}.
In the two limiting cases of $g=0$ and $g\rightarrow \infty$, it can be approximated as
\begin{eqnarray}
\lim_{g \rightarrow 0} \bar{F}(g) &=& - \frac{1}{16\pi} g + \cdots, \nonumber \\
\lim_{g \rightarrow \infty} \bar{F}(g) &=& \frac{1}{8\pi} \left( 1- \frac{\log{\sqrt{g}}}{\sqrt{g}} \right) 
+ \cdots.
\label{limitF}
\end{eqnarray}
Substituting $g= a^2/4M^2$ in $\bar{F}(g)$, one obtains the dependence of the backreaction on the acceleration $a$ 
of the point particle
 and the mass $M$ of the radiation field.

\begin{figure}
\centering
\includegraphics[scale=0.6]{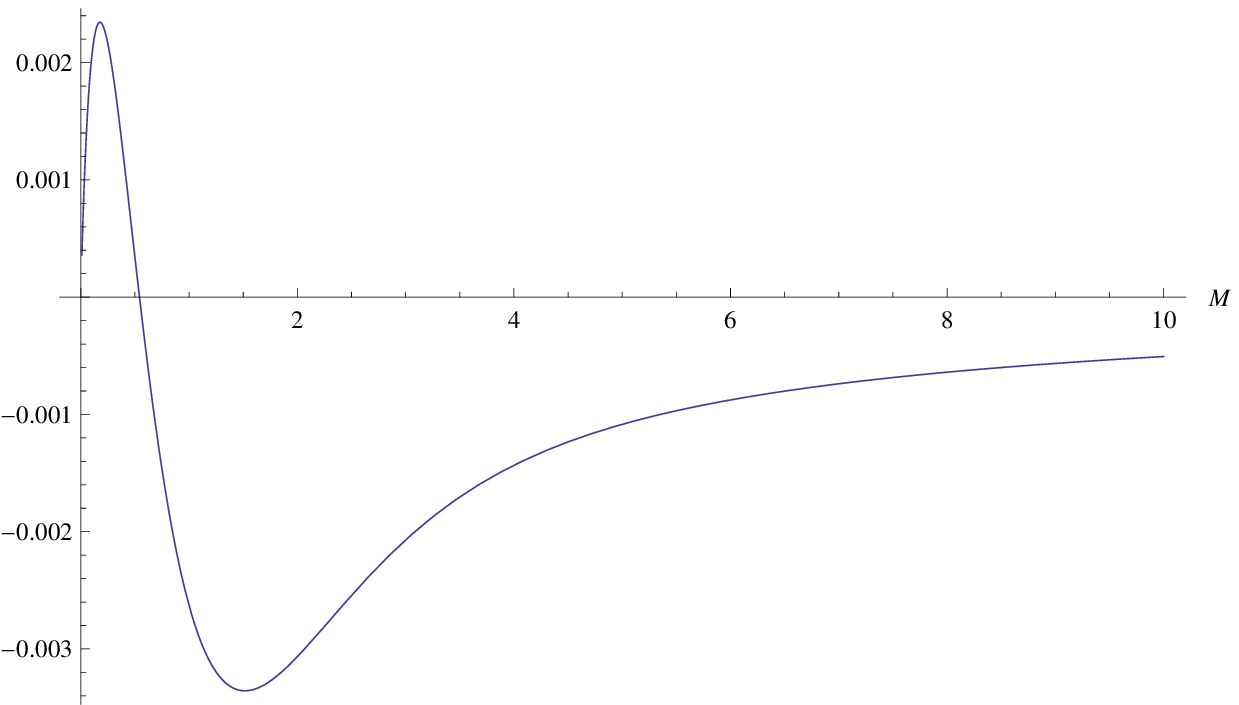}
\caption{$a M\bar{F}(g)$ as a function of $M$, with $a=1$}
\label{uniformm}
%Figure. \ref{uniformm} 
\includegraphics[scale=0.6]{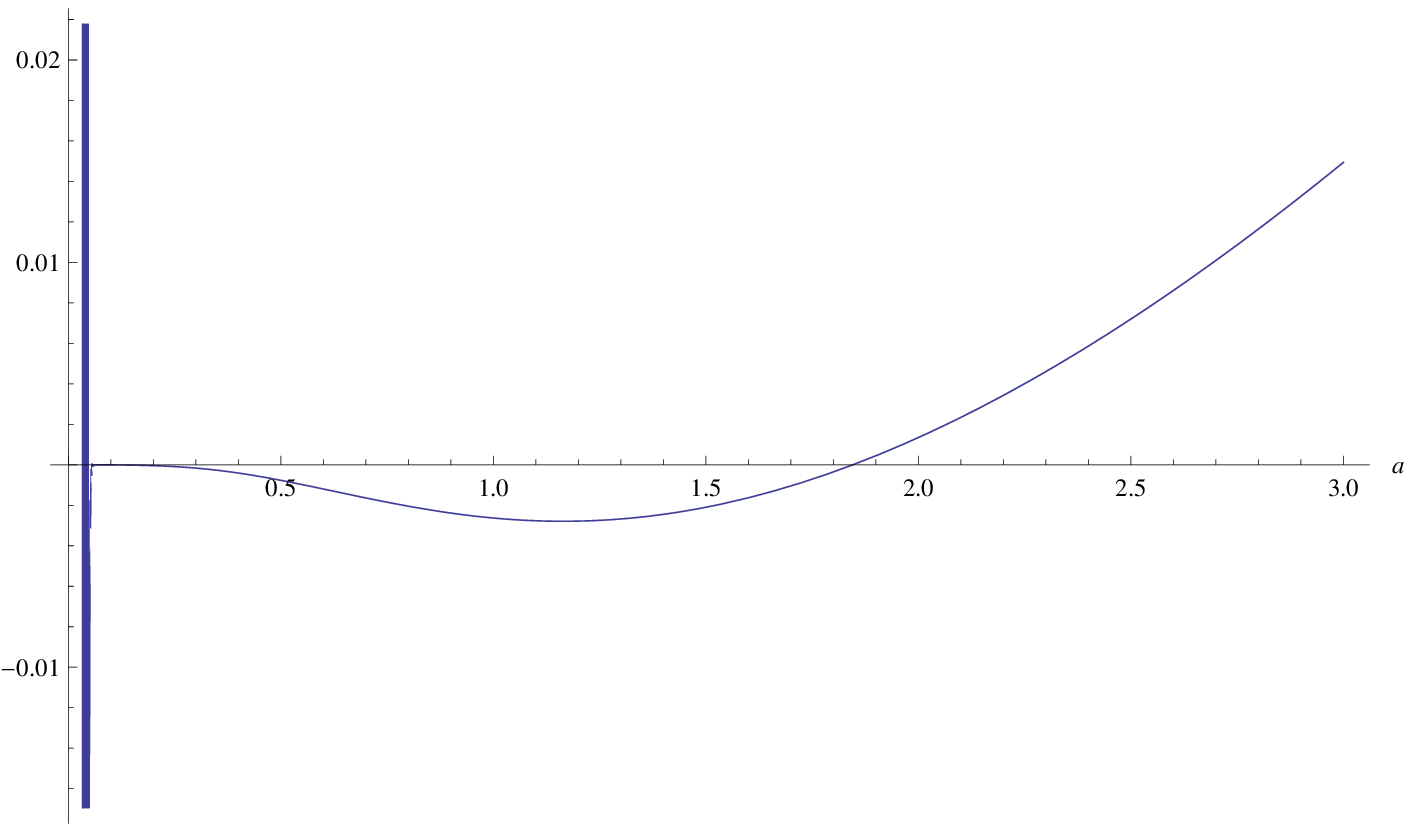}
\caption{$a M\bar{F}(g)$ as a function of $a$, with $M=1$}
\label{uniforma}
%Figure. \ref{uniforma} 
\end{figure}

Let us first look at the behavior of the backreaction $F(a,M)$ 
as a function of $M$ with the acceleration $a$ fixed.
It is plotted in Figure. \ref{uniformm}.
From (\ref{limitF}), one can see that near $M=0$ the backreaction $F(a,M)$ becomes
\begin{eqnarray}
\lim_{M\rightarrow 0} F(a,M)
= \lim_{M\rightarrow 0} e^2 a M \bar{F}(\frac{a^2}{4M^2}) = \frac{e^2 a M}{8\pi} - \frac{e^2 M^2}{4\pi}  \log{M} + \cdots,
\label{FM0}
\end{eqnarray}
Note that $e^2 a M \bar{F}(g)$ vanishes at $M=0$.
However it is not analytic at $M=0$ and contains a logarithmic 
term proportional to $M^2 \log{M}$.
This is the reason why one could not expand the backreaction term with respect to $M$. 
One can also obtain the large mass limit, $M \rightarrow \infty$, as
\begin{eqnarray}
\lim_{M\rightarrow \infty} F(a,M) = 
- \frac{1}{16\pi}\frac{e^2 a^3}{4M} + \cdots.
\label{limitm}
\end{eqnarray}
The backreaction vanishes in the limit as expected.
\footnote{
It is written by an inverse power of $M$, but 
it does not mean that a derivative expansion is valid since an explicit trajectory was used
to derive the result.
}
  
We then fix $M$ and look at the dependence of the backreaction on the acceleration $a$.
The behavior of $F(a,M)= a M \bar{F}(g)$ as a function of $a$ is plotted in Figure. \ref{uniforma}.
In the limit $a\rightarrow \infty$, the back reaction term $F(a,M)$ becomes
\begin{eqnarray}
\lim_{a \rightarrow \infty} F(a,M) = \lim_{a \rightarrow \infty} (\frac{e^2 a M}{8\pi} + e^2 M^2 \log{a}).
\label{loga}
\end{eqnarray}
It is proportional to $a$ in the leading order, and has a correction of $\log{a}$.
The logarithmic factor $\log{a}$ has the same origin as the logarithmic factor $\log{M}$
in (\ref{FM0}).
In the limit $a\rightarrow 0$
 which corresponds to the limit of the point charge staying at the origin, the backreaction 
is suppressed as $a^3$
\begin{eqnarray}
\lim_{a \rightarrow 0} e^2 a M F(g) = - \lim_{a\rightarrow 0} \frac{e^2}{16 \pi}\frac{a^3}{4M} \rightarrow 0
\end{eqnarray}
If we took a different choice of $f(l)$, the coefficient of the term proportional to $a$ did not vanish.
This implies the plausibility of the simplest choice $f(l)=1/2$.
Indeed, if we choose a different $f(l)$, the change of the backreaction term is given by
\begin{eqnarray} 
e^2 a M \int^\infty_0 d t \left( f(t/M) - \frac{1}{2} \right) \frac{J_1(t)}{4\pi t} = e^2 a \delta F(M).
 \label{massrenormalization}
\end{eqnarray}
It is proportional to $a$ multiplied by a function of $M$. 
Though a different choice of $f(l)$ changes the mass renormalization and the leading behavior of the radiation 
reaction in the  $a \rightarrow 0$ or $a\rightarrow \infty$ limit, 
it does not change the logarithmic behavior ($\log{a}$ or $\log{M}$).

Another important feature is the sign of the backreaction $F(a,M)$. 
As shown in figure \ref{uniformm}, $F(a=1,M)$ becomes either negative or positive, and
such a behavior of $F(a,M)$ is not special to $f(l)=1/2$, but also holds for a general choice of $f(l)$
(see the appendix \ref{proof}).
Since the backreaction is given by $F_{self}^\mu = (\ddot{z}^\mu/a) F(a,M)$, 
a negative value of $F(a,M)$ is physically natural since 
it means that 
the backreaction term suppresses a change of the particle motion.

On the contrary, if $F(a,M)$ is positive,  
it enhances the acceleration when the particle is accelerated.
Hence it is an indication of an instability of particle motion
or a breakdown of the framework which we are using.
The positiveness $F(a,M)>0$ for a specific region of the parameters
is generally satisfied for  $a \gg M$. 
It is interesting to investigate a relation with the problem of runaway solutions 
in the massless ALD equation.

%%%%%%%%%%%%%%%%%%%%%%%%%%%%%%%%%%%%%%%%%%%%%%%%%%%%%%%%%% }}}
\subsection{A Scattering Process} %{{{
In the previous section, we have seen a non-analytic behavior of the backreaction for
a uniform acceleration. In this case,  
the point charge is accelerated for an infinitely long time, and the total radiation becomes infinite.
Hence one may suspect that such a   non-analytic behavior is caused by the infinite acceleration.
In this section, we see that a similar non-analytic behavior does arise even when a particle
is accelerated for a finite time interval.

Now we consider a charged particle with a constant four velocity  $v^\mu$ at $\tau<0$. During a finite time
interval $0<\tau<L$, the particle is scattered (or feels an external force) and then after $\tau>L$ it goes away
with a velocity $v'^\mu$.  The trajectory is given by
\begin{eqnarray}
z^\mu = \left\{ 
\begin{array}{c}
v^\mu  \tau, \ \ \ \textrm{for} \ \ \tau<0 \\
q^\mu(\tau), \ \ \ \textrm{for} \ \ 0<\tau<L \\
v'^\mu  \tau + c^\mu, \ \ \ \textrm{for} \ \ \tau > L
\end{array}
\right.
\end{eqnarray}
where $q^\mu(\tau)$ is a function satisfying the boundary conditions
 $q^\mu(0) = 0$ and $q^\mu(L) = c^\mu + v'^\mu L$.

At $\tau < 0$, the backreaction is $0$ since no radiation has been emitted yet.
It is confirmed from the equation (\ref{sform}).
Since $y^\mu = s v^\mu$, the backreaction trivially
 vanishes by the projection operator $P^\mu_\nu = \delta^\mu_\nu - \dot{z}^\mu \dot{z}_\nu$. 
During the time interval
 $0 < \tau < L$, 
 the point particle is scattered by the target (or source of the external field).
 Let us divide the integral (\ref{sform}) into two parts, one with $\int^\tau_0 ds$ and 
the other with $\int^\infty_\tau ds$.
The first part does not cause any non-analytic behavior since 
 the integration range $\int^\tau_0 ds$ is finite
\footnote{
Since the integral is always convergent, we can 
obtain a  power series of $M^n$ by
expanding $M J_1(M l)$ with respect to $M$. 
}.
So we are interested in the second part,
\begin{eqnarray}
F_{\infty}^\mu = - e^2 \int^\infty_\tau ds \ \left(
P^\mu_\nu \frac{d}{d s} \frac{y^\nu}{y^\rho \dot{y}_\rho} + \ddot{z}(\tau)
\right) \frac{M J_1(M l)}{4\pi l}.
\end{eqnarray} 
Using the following relations
\begin{eqnarray}
y^\mu &=& z^\mu (\tau ) - z^\mu (\tau') = q^\mu(\tau) - v^\mu \tau', \ \ \ \ 
\dot{y}^\mu = \frac{d y^\mu}{d s} = v^\mu, \nonumber \\
\sigma &=& y^\mu y_\mu = q^\mu q_\mu + \tau'^2 - 2 q^\mu v_\mu \tau',
\end{eqnarray}
one can rewrite the integral as
\begin{eqnarray}
F^\mu_\infty = -e^2 \int^\infty_\tau ds \left(
\frac{(\delta^\mu_\nu - \dot{q}^\mu \dot{q}_\nu )(v^\nu q^\alpha - q^\nu v^\alpha) v_\alpha}{(v^\rho q_\rho -  \tau')^2} + \ddot{q}^\mu
\right) \frac{M J_1(M l)}{4\pi l}.
\end{eqnarray}
Using the property of $l \sim \tau' \sim s$ at $s \rightarrow \infty$, the integration becomes
\begin{eqnarray}
F^\mu_\infty \rightarrow -e^2 \int^\infty dl \ \left( \ddot{q}^\mu + (\delta^\mu_\nu - \dot{q}^\mu \dot{q}_\nu )(v^\nu q^\alpha - q^\nu v^\alpha) v_\alpha \frac{1}{l^2}(1-2\frac{v^\rho q_\rho + \cdots}{l}) 
\right) \frac{M J_1(M l)}{4\pi l}.
\label{scatteringasymmptotic}
\end{eqnarray} 
The leading power of $l$ at $l\rightarrow \infty$ in the parenthesis is given by the term
proportional to $l^{-3}$.
Due to the property of an integral of the Bessel function (see Appendix \ref{Bessel}),
the integration gives a term proportional to $M^4 \log{M}$.
This term vanishes in the massless limit, but it is non-analytic and we cannot
expand the backreaction term with respect to the mass $M$.
Hence, even for a scattering process where the point particle is
 accelerated only during a finite time interval, 
the backreaction behaves non-analytically at $M=0$.

%%%%%%%%%%%%%%%%%%%%%%%%%%%%%%%%%%%%%%%%%%%%%%%%%%%%%%%%%% }}}
\subsection{A Circular Motion} %{{{
The third example we consider is a circular motion.
When the particle is moving on a circle, the trajectory is given by
\begin{eqnarray}
z^\mu = (\gamma \tau, \rho \cos{\gamma \omega \tau}, \rho \sin{\gamma \omega \tau},0).
\end{eqnarray}
The parameter $\rho$ is the radius of the circular motion and $\rho \omega (< 1)$ is the velocity of the point charge.
These parameters $(\rho,\omega,\gamma)$ must satisfy
\begin{eqnarray}
\gamma = \frac{1}{\sqrt{1-\rho^2\omega^2}},
\end{eqnarray}
in order to hold the gauge condition $\dot{z}^\mu \dot{z}_\mu = 1$.
Since $\dot{z}^\mu$, $\ddot{z}^\mu$ and $\dddot{z}^\mu$ are all independent of each other, the backreaction can be 
generally written in the form
\begin{eqnarray}
F^\mu_{self}(z) = e^2 [F_{m} (0,\cos{\gamma\omega\tau},\sin{\gamma \omega \tau},0) 
+ F_{ALD} (\rho \omega,-\sin{\gamma \omega \tau},\cos{\gamma \omega \tau},0)],
\end{eqnarray}
where the first term is proportional to $-\ddot{z}^\mu$ and the second term is proportional to 
$-(\dddot{z}^\mu +\dot{z}^\mu \ddot{z}^\nu \ddot{z}_\nu)$.
A term proportional to $\dot{z}^\mu$ vanishes due to the projection operator
$P^\mu_\nu$.
The following relations are useful
\begin{eqnarray}
\dddot{z}^\mu + \dot{z}^\mu \ddot{z}^\nu \ddot{z}_\nu = \rho \gamma^5 \omega^3 (-\rho \omega, \sin{\gamma \omega \tau,-\cos{\gamma\omega\tau}},0), \ \ \ \ \ 
\ddot{z}^\nu \ddot{z}_\nu = - \rho^2 \gamma^4 \omega^4, \nonumber \\
\sigma = (z(\tau)^\mu - z(\tau')^\mu)(z(\tau)_\mu - z(\tau')_\mu) 
= \gamma^2 s^2 - 2\rho^2 (1-\cos{\gamma\omega s}).
\end{eqnarray}
%since it is difficult to solve $\sigma$ in terms of $s$, we have evaluate the integration in terms of $s$.
%This makes it difficult to know the analytic properties in this case.
%We are going to show the numerical plot for this case.
The coefficient $F_{m}$ is given by
\begin{eqnarray}
F_{m} = - \int^\infty_0 ds \ 
\rho \left(
\frac{(1+\rho^2 \omega^2 )( \cos{\gamma \omega s} - 1)+\gamma\omega s \sin{\gamma \omega s}}
{(\gamma s - \rho^2 \omega \sin{\gamma \omega s})^2}
+ \gamma^2 \omega^2 (\frac{1}{2}\frac{d l}{d s} - 1)
\right) \frac{M J_1(M l)}{4\pi l},\nonumber \\ 
\label{circularfm}
\end{eqnarray}
and determines the mass renormalization.
%which has an ambiguity from the function $f(l)$.
The other term
 $F_{ALD}$ gives a coefficient of the ALD term
\begin{eqnarray}
F_{ALD} = 
- \int^\infty_0 ds \ 
\rho \frac{\gamma \omega s \cos{\gamma \omega s} - \sin{\gamma \omega s}}
{(\gamma s - \rho^2 \omega \sin{\gamma \omega s})^2} \frac{M J_1(M l)}{4\pi l}
- \frac{\rho \gamma^5 \omega^3}{12\pi}.
\label{circularfald}
\end{eqnarray}
The second term $-(\rho \gamma^5 \omega^3/12\pi)$ is equal to the ALD force when the radiation field
is massless while 
the first term gives a correction in the massive case and comes from the non-local part of the Green function.

\begin{figure}
\centering
\includegraphics[scale=0.6]{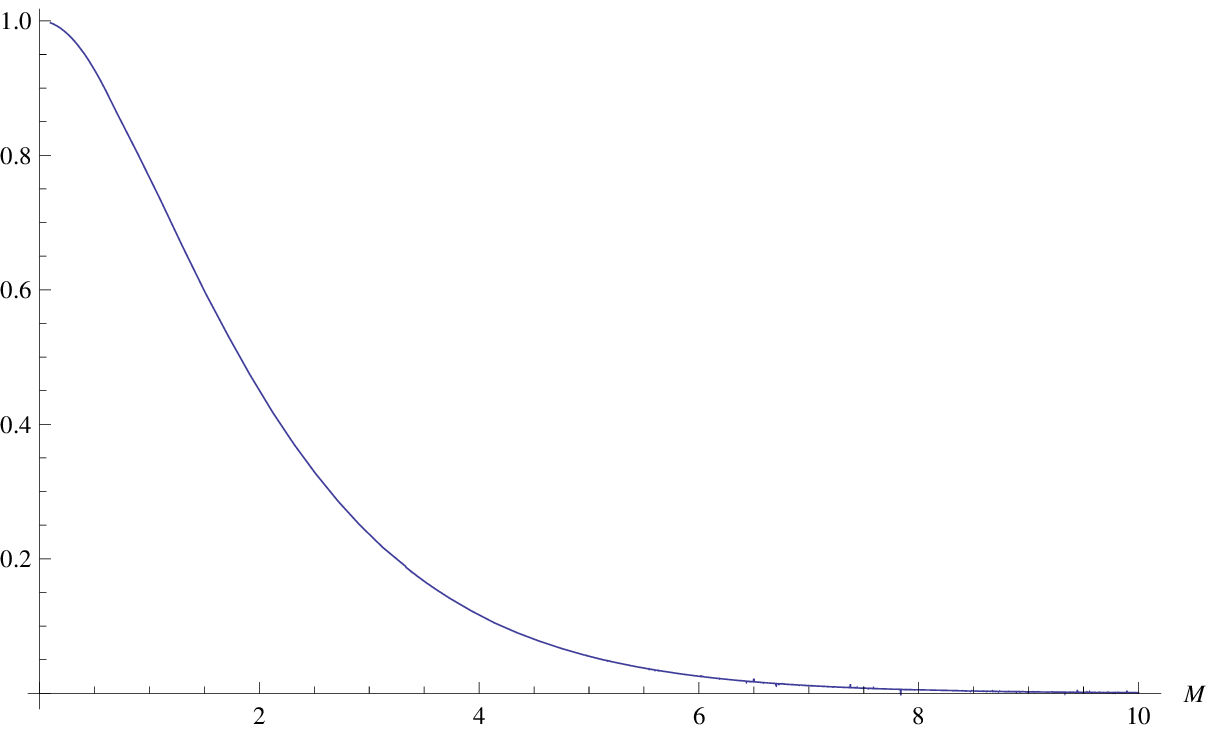}
\caption{$F_{ALD}/F_{ALD}^{massless}=F_{ALD}/(- \rho \gamma^5 \omega^3 / 12\pi)$ as a function of $M$, with $\rho=1$ and $\omega = 0.8$}
\label{circularm}
%Figure. \ref{circularm} 
\includegraphics[scale=0.6]{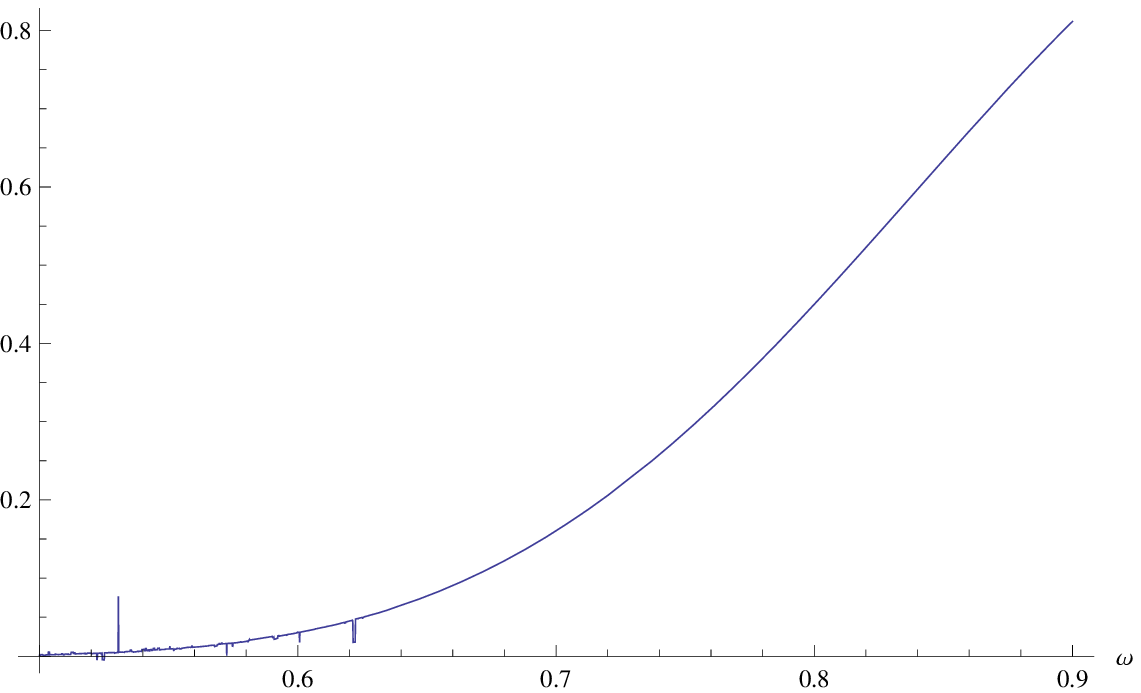}
\caption{$F_{ALD}/F_{ALD}^{massless}$ as a function of $\omega$ with $\rho=1$, $M = 2$}
\label{circularw}
%Figure. \ref{circularw} 
\end{figure}  

In order to see  the dependence of the backreaction on $M$, we first note that 
 $l \sim \gamma s$ at $s\rightarrow \infty$. Then 
both of the integrands of $F_{ALD}$ and $F_{m}$ are
 proportional to $\sin{\omega l}$ or $\cos{\omega l}$ at $l \rightarrow \infty$
\footnote{ 
There is also a non-oscillating term 
 $ \sim l^{-2}$ in $F_m$. But as we see in the appendix \ref{Bessel},
such a term with an even power of $l$ does not produce a logarithmic behavior
 at $M=0$. 
}.  Thus there are no non-analytic terms such as $\log{M}$(see appendix \ref{Bessel} for details)
in the case of the circular motion.

Finally let us  see how the radiation reaction changes as a function of $M$.
In Figure. \ref{circularm},  the ratio of $F_{ALD}$ to the radiation reaction in the massless case is plotted.
It becomes $1$ at $M \rightarrow 0$ and decreases as the mass of the radiation field $M$
increases. Eventually it  vanishes at $M \rightarrow \infty$. This result is consistent with our
expectation that the radiation reaction is suppressed by the effect of the mass of the radiation field.
In Figure. \ref{circularw} we plot the same ratio as a function of $\omega$. 
The ratio becomes $1$ in the relativistic limit $\rho \omega \rightarrow 1$.
When the frequency $\omega$ decreases and the particle's motion becomes 
nonrelativistic, the radiation reaction is more suppressed in the massive case.
This can be naturally expected since in the nonrelativistic region with a small $\omega$
the emission is highly suppressed by the mass of the radiation field.

%%%%%%%%%%%%%%%%%%%%%%%%%%%%%%%%%%%%%%%%%%%%%%%%%%%%%%%%%
%%%%%%%%%%%%%%%%%%%%%%%%%%%%%%%%%%%%%%%%%%%%%%%%%%%%%%%%%% }}}
\section{Conclusions } %{{{
\setcounter{equation}{0} 
In this paper, we investigated the radiation reaction of a charged particle interacting with a massive scalar field.
We first obtained a massive analog of the ALD equation.
The most important observation is that the massive ALD equation cannot be written as a local equation
with higher derivative terms. 
The coefficients of higher derivative terms are divergent and
the derivative expansion of the radiation reaction term  is invalid.
Only in the massless limit the ALD equation becomes local. 
 This sounds strange since the massless limit is more sensitive
to the infrared effect and the nonlocal effect seem to become more important than a massive case.
Technically speaking the locality arises in the massless ALD equation
 because the retarded Green function of a massless field has a support
only on the light cone and is proportional to $\delta(\sigma)$.
On the contrary, the Green function of a massive field is distributed around the light cone
and thus
the radiation reaction term becomes nonlocal which is written 
only as an integral form. 

We also studied  the nonanalytic behavior of the radiation reaction term in various situations.
First we showed that it has a non-analytic behavior as 
$M^p \log M^2$ at $M \rightarrow 0$, where $p$ is a positive integer
depending on the details of the trajectory of the particle.
Such non-analytic behavior generally appears at $M \rightarrow 0$
even when the particle is accelerated during a finite time interval.
In appendix \ref{Bessel} we studied how such a non-analytic behavior with 
$\log M$ appears by looking at various integrals of the Bessel function.
We  also evaluated the backreaction in
specific motions, a uniform acceleration and a circular motion.
In both cases, we evaluated the backreaction (radiation reaction) as a function of 
the mass $M$ and showed that it is suppressed when the mass becomes large.
It is consistent with our physical intuition that the emission  is suppressed
in the massive case and accordingly the backreaction is reduced.

%
%For future work, it is worth to investigate further about the singular behavior of the backreaction like $\log{M}$, and separate the regular part from such terms contain $\log{M}$.
%This sounds natural since the singular behavior comes from the integral at infinity.
%It is also interesting to use the same way to investigate the non-local effect of the backreaction for other cases, for example, the backreaction for space-time dimensions other than $D=4$, the backreaction at curved space time and the backreaction due to the radiation of the gravitational wave etc.
%Another direction is to find the correspondence to the quantum field theory.
%As we have mentioned, one need to consider the mass renormalization carefully to keep the consistency.
%And one can not completely determine the mass renormalization for massive case, while the issue is unique for the massless case.

Finally we would like to comment on a possible resolution to the runaway solutions in the massless ALD equation.
As we saw at the end of Section \ref{uniformlyacceleration},  
the positive region of the backreaction $F(a,M)$ suggests that
 an instability will occur in the region $a \gg M$. 
It may be related to the pathological behavior of the massless ALD equation.
In the massive case, when a particle is accelerated from a nonrelativistic region,
the backreaction term $F(a,M)$ is negative and suppresses the acceleration. Hence
it tends to avoid the runaway type solution. 
It is interesting to
investigate that if one can avoid the pathology of the runaway
solution by introducing a small mass of the radiation field and then
taking a massless limit.

%For the massless case, one can find a example of such approach at \cite{Higuchi:2006xk}, where the  cancellation of the infinity and the mass renormalization is shown in a form which closely related to the classical case.
%However, for massless case one just derived the ALD equation, but for massive case, one may find something new. 
%As mentioned at the end of Section \ref{uniformlyacceleration}, it is suggested that there exist some instability at the region $a >> M$.
%It is interesting to find if there are some physical origin for such kind of instability.
%Since for the massless case $M=0$, then it is always satisfied that $a>>M$.
%This may shed some light on the problem of runaway solution or pre-acceleration.
% 

%%%%%%%%%%%%%%%%%%%%%%%%%%%%%%%%%%%%%%%%%%%%%%%%%%%%%
%%%%%%%%%%%%%%%%%%%%%%%%%%%%%%%%%%%%%%%%%%%%%%%%%%%%% 
%%%%%%%%%%%%%%%%%%%%%%%%%%%%%%%%% }}}
\section*{Acknowledgments} %{{{
We would like to thank S.V. Bulanov and K. Seto for discussions.
The research by S.I. is supported in part by Grant-in-Aid for Scientific Research (19540316) from MEXT, Japan.
We are also supported in part by "the Center for the Promotion of Integrated Sciences (CPIS) "  of Sokendai.

%%%%%%%%%%%%%%%%%%%%%%%%%%%%%%%%%%

%%%%%%%%%%%%%%%%%%%%%%%%%%%%%%%%%
\appendix %}}}
\section{Some properties of $J_1(x)$} %{{{
\label{Bessel}
The Bessel function $J_1(x)$ is given by the series 
\begin{eqnarray}
J_1(x) = \sum^{\infty}_{n=0} \frac{(-1)^n (x/2)^{2n+1}}{n!\Gamma(n+2)},
\end{eqnarray}
which is an odd function of $x$. It is regular on the whole complex plane except at infinity. 
The asymptotic behavior of $J_1(x)$ is given by
\begin{eqnarray}
J_1(x) \sim \sqrt{\frac{2}{\pi x}} \cos(x - \frac{3\pi}{4}) + o(x^{-3/2}).
\label{asymptoticJ}
\end{eqnarray}
Though it falls off  very slowly as $1/\sqrt{x}$ at infinity, 
 the oscillating factor makes the following integral  convergent
\begin{eqnarray}
\int^\infty_0 d x \ J_1(s) = 1.
\end{eqnarray}
In this appendix, we focus on the behavior of the function 
\begin{eqnarray}
I(M,l) = \frac{M J_1(M l)}{l}.
\end{eqnarray}
and its integrals 
in the two limiting situations,  $M \rightarrow \infty$ and $M \rightarrow 0$.

%%%%%%%%%%%%%%%%%%%%%%%%%%%%%%%%%
\subsection{$M \rightarrow \infty$ limit}
We first show that the function $I(M,l)$ has the following property
\begin{eqnarray}
\lim_{M \rightarrow \infty} I(M,l) = 2 \delta(l^2).
\label{delta}
\end{eqnarray}
In order to show this, we introduce a continuous and regular function $g(l)$ 
in the region $0 \le l < \infty$ with the boundary condition 
$\lim_{l \rightarrow \infty} g(l) =0$.
Then the following integral becomes
\begin{eqnarray}
\lim_{M\rightarrow \infty} \int^\infty_0 d(l^2) \ I(M,l) g(l^2) &=& 
\lim_{M\rightarrow \infty} \int^\infty_0 dl \ 2 M J_1(M l) g(l^2) \nonumber \\
&=& \lim_{M\rightarrow \infty} 2 \int^\infty_0 ds J_1(s) g(s^2/M^2) 
 \nonumber \\
 &=& 2 \int^\infty_0 ds J_1(s) g(0) = 2 g(0).
 \end{eqnarray} 
In the last equality we changed the ordering of $\lim_{M\rightarrow \infty}$ and $\int^\infty_0 ds$.
If it is justified, the equality (\ref{delta}) is proved.
To justify it, one needs to show that
$$\lim_{M \rightarrow \infty}J_1(s) g(s^2/M^2) = J_1(s) g(0)$$
is  uniformly  convergent at $0 \leq s < \infty$.
%Generally, it isn't possible to find a $s$ independent number $N_\epsilon$ which satisfies
%\begin{eqnarray}
%|g(s^2/N^2_\epsilon) - g(0)| < \epsilon
%\end{eqnarray}
%for all of $s$, 
Namely we need to show that for any $\epsilon$ there exists $N_\epsilon$ such that
the inequality 
\begin{eqnarray}
|J_1(s) \{ g(s^2/N^2_\epsilon) - g(0) \} |<\epsilon,
\label{uniformconvergence}
\end{eqnarray}
is satisfied for all $0 \le s <\infty$.

Such  $N_\epsilon$ can be found as follows.
Let us denote the maximum value of $|g(s)-g(0)|$ for $0 < s < \infty$ by $\delta g_{max}$.
Then, since $J_1(s)$ behaves like $1/\sqrt{s}$ at large $s$, 
there exists $S_\epsilon$ such that $|J_1(s) \delta g_{max}|<\epsilon$ is satisfied for $s \ge S_\epsilon.$
On the other hand, for the region $s \le S_\epsilon$,
we can always find $N_{\epsilon}$
such that $|g(s^2/N^2_{N_{\epsilon}}) - g(0)|<\epsilon$ 
since $g(l)$ is  continuous at $l=0$.
To summarize,  for the region $s \ge S_\epsilon$, 
 $|J_1(s)\{ g(s^2/N^2_\epsilon) - g(0) \}| < |J_1(s) \delta g_{max}|<\epsilon$ is satisfied
and  for the region $s \le S_\epsilon$, 
$|J_1(s)\{(g(s^2/N^2_\epsilon) - g(0)\}| < |g(s^2/N^2_\epsilon) - g(0)| < \epsilon$.
Thus the uniform convergence (\ref{uniformconvergence}) is proved.

%%%%%%%%%%%%%%%%%%%%%%%%%%%%%%%%%
\subsection{Massless limit: $M \rightarrow 0$}
 
We then  investigate the behavior of the integration
\begin{eqnarray}
Q(M) = \int^\infty_0 d l \ I(M,l) g(l),
\end{eqnarray}
in the massless limit $M \rightarrow 0$.
One can similarly prove that
\begin{eqnarray}
\lim_{M\rightarrow 0} I(M,l)g(l) = 0
\end{eqnarray}
is uniformly convergent at $0\leq l<\infty$.
We can exchange the ordering of the limit and the integral and show that
\begin{eqnarray}
Q(0) = \lim_{M\rightarrow 0} Q(M) = 0.
\end{eqnarray} 
Hence
the non-local term like (\ref{nonlocalterm}) vanishes  in the massless limit
and it is reduced to the massless ALD equation.

However, the analytic behavior of $Q(M)$ is not so simple
since  logarithmic terms like $M^2 \log{M}$ are expected to appear
around $M=0$.
In the following, we show how to derive such a non-analytic term of $Q(M)$.
First note that the integrand $I(M,l)$ is analytic 
for both parameters, $l$ and $M$. Then an integration over a finite range
\begin{eqnarray}
Q(M;L) = \int^{\Delta}_0 d l \ I(M,l) g(l),
\end{eqnarray}
is also an analytic function of $M$.
Non-analytic behavior like $\log{M}$ appears only when we take the integral region infinite
 $\Delta \rightarrow \infty$.

Let $g(l)$  be an analytic function at finite $l$ which falls off as $l^{-n}$ at infinity.
Here $n$ is a non-negative integer. 
We encountered two examples in the analysis of the radiation reaction.
In the case of  the uniform acceleration (\ref{uniformmaster}), 
an integral with $g(l) \sim l^{-1}$ appears.
In a scattering process (\ref{scatteringasymmptotic}), $g(l)$ contains a term proportional to $l^{-3}$. 
In order to see the non-analytic behavior of these integrals, 
we divide the integral $Q(M)$ into a regular part $Q(M;L)$ and 
a (possibly) singular part $Q(M;\infty)$ as 
\begin{eqnarray}
Q(M) = Q(M;L) + Q(M;\infty) = \int^L_0 d l \ I(M,l) g(l) + \int^\infty_L d l \ I(M,l) g(l), 
\end{eqnarray} 
where $L$ is chosen to be large enough 
so that one can make the approximation $g(l)=\alpha l^{-n}$.
% (to be large enough that the expansion of $g(l)$ around $l=\infty$ is valid).
By changing the integration variable to $s = ML$, the second part becomes 
\begin{eqnarray}
Q(M,\infty) = M \int^\infty_{M L} d s \ \frac{J_1(s)}{s} g(s/M) 
\approx \alpha M^{n+1} \int^\infty_{M L} d s \ \frac{J_1(s)}{s^{n+1}}.
\end{eqnarray}
The  approximation in the second equality 
is valid even in the massless limit because $s/M$ is always larger than $L$ and 
$g(s/M) \sim \alpha (s/M)^{-n}$ is a good approximation.
The $M$ dependence  only comes from the integration around $s = M L$.
For a small $M$, one can expand $J(s) \approx s/2 - (s)^3/16 + \cdots$ near $s = M L$, and $Q(M,\infty)$ becomes
\begin{eqnarray}
Q(M,\infty) &\approx & \alpha M^{s+1} \int_{M L} ds \ \sum^{\infty}_{k=0} \frac{(-1)^k s^{2k-n}}{2^{2k+1} k! (k+1)!} 
\nonumber \\
&=& \alpha M^{2k+2} \sum^{\infty}_{k=0, k\neq \frac{n-1}{2}} \frac{(-1)^{k+1}}{2^{2k+1} k! (k+1)!} \left(
\frac{L^{2k-n+1}}{2k-n+1} 
+ \delta_{k,\frac{n-1}{2}} \log{ML}
\right).
\end{eqnarray}
It is a regular function for an even integer $n$. For an odd integer $n$,
the logarithmic term $\log{M}$ arises.
The coefficient of the logarithmic term $\log{M}$ is given by
  \begin{eqnarray}
\alpha M^{n+1} 
\frac{(-1)^{\frac{n+1}{2}}}{2^n \left( \frac{n-1}{2} \right) ! \left( \frac{n+1}{2} \right) !}.
\end{eqnarray}
It is determined by the behavior of $g(l) \sim \alpha l^{-n}$ at infinity  and the behavior of 
the Bessel function $J_1(s)$ near $s=0$.
As an example, due to the asymptotic behavior  (\ref{limitm}), the integral (\ref{uniformmaster}) at $a=1$ is 
reduced to the above integral $Q(M)$ with $n=1$ and $\alpha = - e^2/2\pi$, and
 the logarithmic dependence in (\ref{loga}) is reproduced.

When $g(l)$ is an oscillating function, the behavior of the integral  becomes different.
It is the case corresponding to the circular motion in (\ref{circularfm}) and (\ref{circularfald}).
In this case, $g(l)$ contains oscillating factors like $\sin{k l}$ at $l \rightarrow \infty$.
Suppose that $g(l)$ behaves like $g(l) \sim l^{-n} \sin{k l}$ for $l \rightarrow \infty$.
If we divide the integral $Q(M)$ into $Q(M,L)$ and $Q(M,\infty)$ as above,
a non-analytic behavior, if exists, must come from the second part $Q(M,\infty)$.
By changing the integration variable to $s = M L$, it becomes
\begin{eqnarray}
Q(M,\infty ) \approx \alpha M^{n+1} \int^\infty_{M L} ds \ \frac{J_1(s) \sin{(k s/M)}}{s^{n+1}}.
\end{eqnarray}
Since $ks/M > kL \gg 1$, the term
$\sin{(k s/M)}$ is oscillating rapidly.
Since $J_1(s) \sim s$ for small $s$, $M$-dependence of $Q(M,\infty)$ is determined by
the integration,
\begin{eqnarray}
\int^\infty_{M L}  \frac{\sin{k s/M}}{s^n} ds = M^{n+1} \int^{\infty}_{L} \frac{\sin{k t}}{t^n} d t,
\end{eqnarray}
which is a power function of $M$.
Thus no non-analytic behavior appears when 
 $g(l)$ is an  oscillating function at infinity. 
%%%%%%%%%%%%%%%%%%%%%%%%%%%%%%%%%%%%%%%
\section{Existence of a positive region of $F(a,M)$ }
\label{proof}
In this appendix we prove that, in the uniform acceleration,  
whatever $f(l)$ we choose
 the backreaction must  take a positive value for a sufficiently large $g=a^2/4M^2$.
This can be shown as follows.
If we take a different choice of $f(l)$ from the simplest choice $f(l)=1/2$,
the backreaction  changes as in (\ref{massrenormalization}).
This change of the backreaction must satisfy the condition
\begin{eqnarray}
\lim_{M\rightarrow \infty} \delta F(M) = 0
\end{eqnarray}
since it should vanish in the large $M$ limit.
Hence, for any small $\epsilon$, there exists $M_\epsilon$ that satisfies $|\delta F(M)| < \epsilon$ for $M>M_\epsilon$.
On the other hand, as shown in Figure. \ref{uniformg},
the function $\bar{F}(g)$ is positive for $g>1$ and grows as a function of $g$ and there exists $g_\epsilon$
such that $\bar{F}(g) > \epsilon$ for $g>g_\epsilon$.
For a fixed $M>M_\epsilon$, we can always take $a$ sufficiently large so that $g=a^2/4 M^2 >g_\epsilon$.
Then in such values of $a$ and $M$, the backreaction becomes positive;
\begin{eqnarray}
\frac{F(a_l,M_l)}{e^2 a_l M_l} = \delta F(M_l) + \bar{F}(\frac{a^2_l}{4M^2_l}) > 0.
\end{eqnarray}

%%%%%%%%%%%%%%%%%%%%%%%%%%%%%%%%% }}}

\end{document}